\documentclass[fleqn,twoside]{article}
\usepackage{espcrc2}

\usepackage{graphicx}
\usepackage{epsfig}
\usepackage[figuresright]{rotating}

\newcommand{\AmS}{{\protect\the\textfont2
  A\kern-.1667em\lower.5ex\hbox{M}\kern-.125emS}}

\title{Low-Energy Hadron Production Data and 
Current Status of CERN Measurements\thanks{
Based on a talk given at the XIII International Symposium on Very High
Energy Cosmic Ray Interactions, Pylos, Greece, 6-12 Sep 2004.}
}

\author{
  Giles Barr\address{University of Oxford, Department of Physics,\\
    Denys Wilkinson Building, Keble Road, Oxford, OX1 3RH, United Kingdom}
and 
  Ralph Engel\addressmark[FZK]\address{Forschungszentrum Karlsruhe,
    Institut f\"ur Kernphysik, 76021 Karlsruhe, Germany}
}

\begin{document}

\begin{abstract}
Data on low-energy hadron production in collisions of nucleons, pions
and kaons with light nuclei are needed for many astrophysical and
accelerator applications. Modern simulations have reached a level 
of accuracy that
the lack of detailed understanding of hadron production processes has
become one of the most important limitations to further improvements.
After giving some examples of hadroproduction processes in astrophysics
and neutrino experiments we briefly review existing fixed-target data 
on light nuclei. Preliminary results and prospects of current CERN 
measurements (HARP, NA49) are discussed.
\end{abstract}

\maketitle


\subsection*{Introduction}

The increase in computational power in recent decades has allowed some
very precise experimental measurements to be made.  In particle
physics, detector effects have been simulated using the Monte Carlo
technique to the sub-percent level to obtain accurate corrections to
measurements.  Effects such as particle decay dynamics,
electromagnetic shower propagation and energy loss are understood to a
sufficient accuracy to allow these precise corrections to be
determined.  The interactions of hadrons, on the other hand reveal a
considerable gap between the accuracy available in simulation programs
and what is desired for certain applications.  The reason for this is
because the underlying theory of the strong interaction, Quantum
Chromodynamics (QCD), is computationally difficult except for processes
with large momentum transfer where perturbation theory can be applied.

The authors of simulation programs for hadron interactions have
adopted either an empirical approach (insert tables and parametrizations 
of accelerator data into the computer code, for example,
\cite{Fesefeldt85a,Engel:2003bv}) or a semi-empirical 
approach (design
models which are motivated from a theoretical point of view and then
use accelerator data to tune the free parameters in the model to make
the simulation fit the data, for example, \cite{Fasso01a,Bleicher99a}).  
These procedures both suffer from a
lack of available hadron production data.  As described below, good
data is only available in a small fraction of the kinematically
allowed and physically important region and for a
small sample of target nuclei.  Model builders
are frequently faced with a choice in how to perform the extrapolation
into uncharted parts of the phase space.  

This article begins by describing some applications drawn from both
cosmic ray physics and neutrino physics where more accurate
simulations of hadron production would considerably 
improve the interpretation of
existing data in terms of fundamental parameters or observables.
We continue with a review of the current state of hadron production
measurements; the authors' brief was to cover CERN experiments,
although other experiments are also mentioned.


\subsection*{Muon production in extensive air showers}

\begin{figure}[htb!]
\begin{center}
\epsfig{file=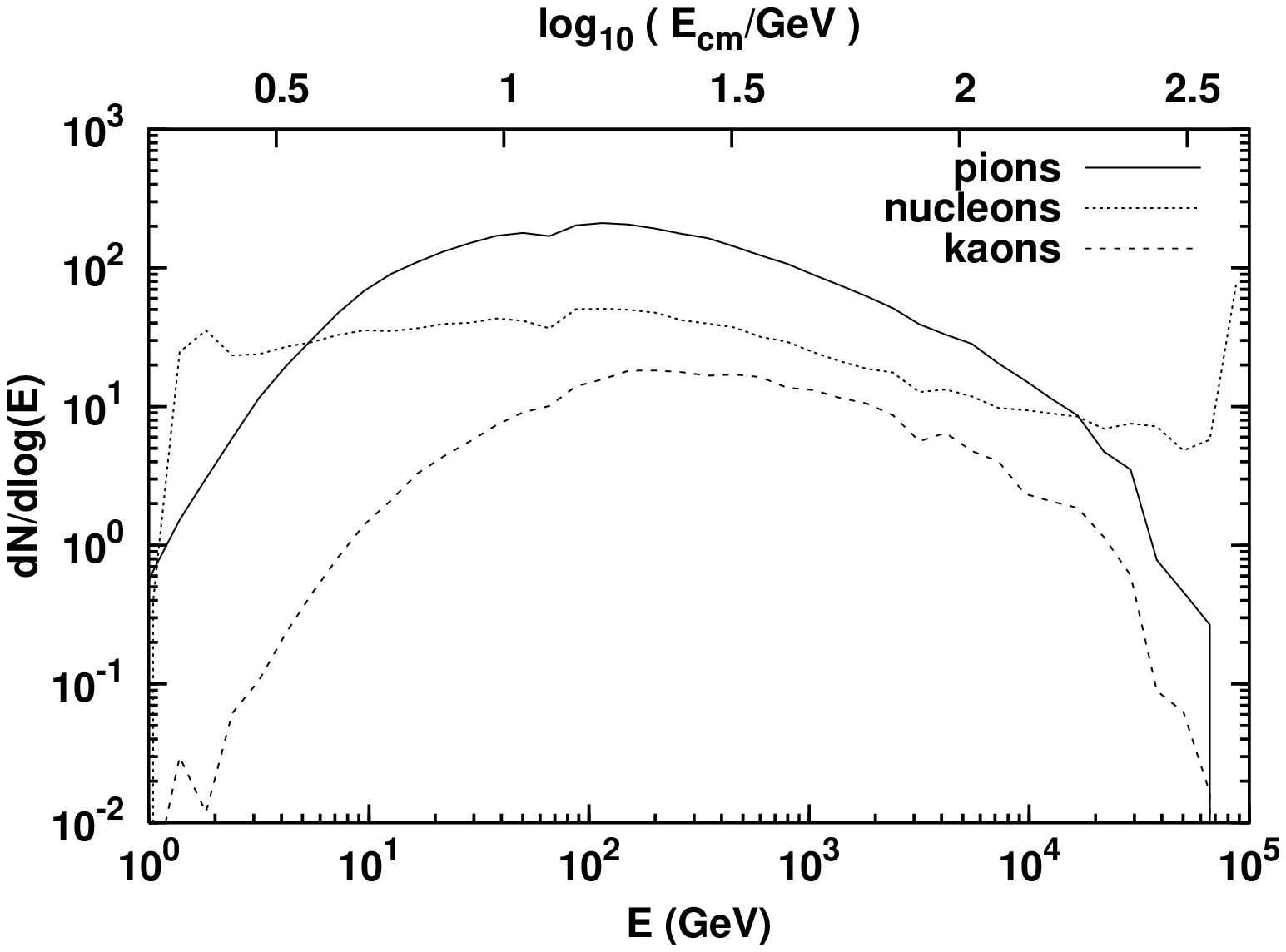,width=7.5cm}
\epsfig{file=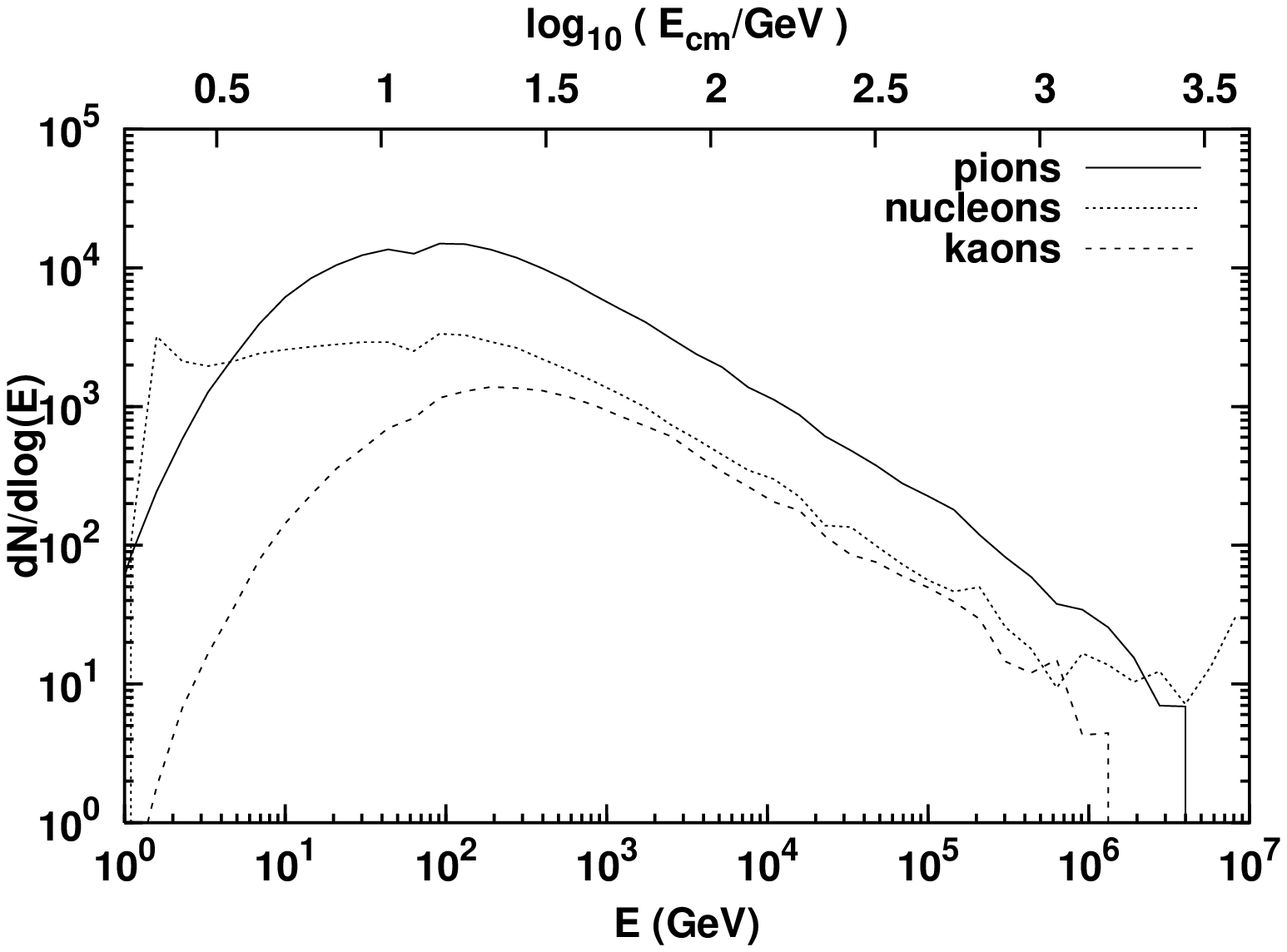,width=7.5cm}
\end{center}
\caption{Energies and initiating particle type 
of interactions which lead to muon
production in showers of 
$E_0=10^{14}$\,eV (top) and $E_0=10^{16}$\,eV (bottom). The upper
horizontal axes show the equivalent hadron-nucleon c.m. energy.
}
\label{figeas}
\end{figure}
Extensive air showers (EAS) are produced by high energy cosmic rays
colliding with air nuclei in the upper atmosphere.  The accurate
simulation of the resulting cascades may be used to predict the
response (e.g. the distribution of muons on the Earth's surface).
These simulations can then be used as a detector calibration; to
deduce the characteristics of the original cosmic rays (e.g.\ the
primary energy) based on the measurements in a detector.  While the
original cosmic ray might be of an energy which exceeds by far the energy
available in accelerator based experiments, the showers contain a
considerable number of lower energy interactions. The number of muons in
EAS at detector level, often used together with the number of
electrons to determine the primary energy and particle type of the
shower, is particularly sensitive to hadron production processes in the
energy range from a few tens of GeV to several TeV \cite{Engel99c}.

Figure~\ref{figeas} shows the energy distribution of those hadronic
hadronic interactions in EAS that produce observable muons.
For each muon arriving at detector level (1032 g/cm$^2$) with an energy
$E_\mu > 250$\,MeV, one entry is made at the energy of hadron $h_{\rm GM}$
\begin{equation}
h_{\rm GM} + {\rm air} \rightarrow h_{\rm M}+X;
\hspace*{1cm} h_{\rm M} \rightarrow \mu + X^\prime ,
\nonumber
\end{equation}
where $h_{\rm GM}$ and $h_{\rm M}$ can be considered as
the muon's grandmother and mother particles, respectively.
The simulations are done with CORSIKA \cite{Heck98a} using
QGSJET \cite{Kalmykov97a} as high-energy hadronic interaction model and 
GHEISHA \cite{Fesefeldt85a} for interactions below 80\,GeV.
The curves illustrate the relative importance of different 
hadron interactions at various
energies. Due to the competition between interaction and decay
processes, the most probable energy is always between 100 to 200\,GeV 
for charged pions and somewhat higher for charged kaons, almost
independent of the primary EAS energy.
This energy range of interactions relevant to
low-energy muon production in EAS is in the reach of fixed-target accelerator 
experiments. Uncertainties in the prediction of muon numbers due to
modeling of low-energy interactions are
presented in Refs.~\cite{Drescher:2002vp,Heck:2003br}. The uncertainties
increase with lateral distance from the shower core and exceed, for
example, 20\% at 2\,km distance and $10^{19}$\,eV.
Therefore, a considerable gain in the quality of EAS
simulations can be expected with new hadron production measurements.


\subsection*{Neutrino production in the atmosphere}

Atmospheric neutrinos have become an exciting study as the Earth, it
turns out, is just about the most convenient size to see neutrino
oscillations~\cite{Fukuda:1998mi} by comparing the zenith angle and energy
distributions with calculations (for a review, see \cite{Gaisser:2002jj}).
Even with an unsophisticated calculation, it is clear that a deficit
of upward going muon neutrinos exists which is interpreted as the
oscillation of $\nu_\mu \rightarrow \nu_\tau$.

\begin{figure}[htb!]
\centerline{\epsfig{file=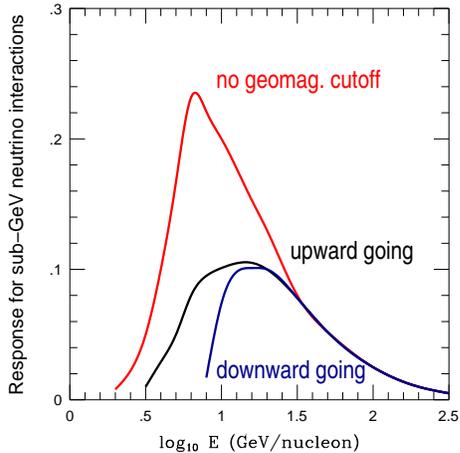,width=6cm}}
\caption{
Response for sub-GeV muon-like events
in SuperKamiokande to the energy of the primary cosmic
ray nucleons (from \protect\cite{Gaisser:1998hb}). Shown are simulations for
solar minimum and no geomagnetic cutoff,
events from lower hemisphere (upward going leptons), and 
events from upper hemisphere (downward going leptons).
}
\label{response}
\end{figure}

Cosmic rays with energies extending down from EAS energies to a few
GeV collide with the atmosphere.  Neutrinos are produced in the decay
of muons, pions and kaons in the resulting cascades.  The neutrinos
which are observed as contained events in underground detectors
($E_\nu \sim 1$~GeV) are produced mostly by the decay of pions from
primaries in the range 2--200~GeV, see Fig~\ref{response}.  
Higher energy neutrinos can be
observed as upward going muons crossing the detector and are produced
by primaries with energies up to $\sim10$~TeV.  At these higher
energies, kaon production becomes extremely important due to their
shorter lifetime than pions (which makes them more likely to decay
than interact).  Additionally, the energy available in the center of 
mass is much larger for $K \rightarrow\mu\nu$ than for 
$\pi\rightarrow\mu\nu$ and the steeply falling cosmic ray spectrum 
increases the importance of these kaon decays.   
With the recent improvements in primary flux
measurements, the lack of comprehensive hadron production measurements
over a more extensive fraction of the phase space is currently
dominating the uncertainty of the atmospheric neutrino fluxes.


\subsection*{Accelerator neutrino beams}

Accelerator neutrino experiments have been used to extensively study
the interactions of neutrinos including the discovery of neutral
currents, precision electroweak measurements and the search and study
of oscillations.  The accelerator beams are produced in a similar way
to the atmospheric neutrinos described above.  Protons from the
accelerator hit a target producing pions.  The neutrinos produced in
the decay of the pions (mostly $\nu_\mu$) are used for the experiment
and the beam is stopped before many of the muons can decay.  The
neutrinos which are produced have a broad energy spectrum and are
mostly $\nu_\mu$ however there is a contamination of $\nu_e$,
$\overline{\nu_\mu}$ and $\overline{\nu_e}$ from muon and kaon decay.
Precision measurements rely on an accurate knowledge of the neutrino
energy spectrum and this is obtained from simulation with similar
limitations from hadron production as for the applications described
above.  Many of the hadron production experiments have been performed
by neutrino physicists with a specific goal of understanding their
neutrino beam.  For a review, see Ref.~\cite{Catanesi04a}.

A radical new way of producing very intense accelerator neutrinos has
been proposed (a neutrino factory) in which a beam is first produced
in the way described above from a low ($\sim 2$--10~GeV) energy proton
beam.  After the pions have been allowed to decay, the muons are
collected and then inserted into a further accelerator where they are
accelerated to energies $E_\mu \sim 50$~GeV.  These muons are then
circulated in a storage ring with long straight sections pointing
towards several detector sites.  The decay of the muons in the
straight sections produces an intense source of neutrinos of known
composition and spectrum.  The gain in the concept is that power is
applied by the accelerator directly to the muons.  There are many
difficulties to overcome, such as the construction of the target
region, and the method of inserting the muons into the accelerator.
It is possible in the further future that the concept may permit a
muon collider to be constructed.  One of the purposes of the HARP
experiment, to be described below, is to investigate the hadron
production in order to optimize the efficiency of a neutrino factory
target.

To summarize, there is considerable interest in the improvement of
hadron production models for many applications.


\subsection*{Currently available data}

\begin{figure}[htb!]
\centerline{\epsfig{file=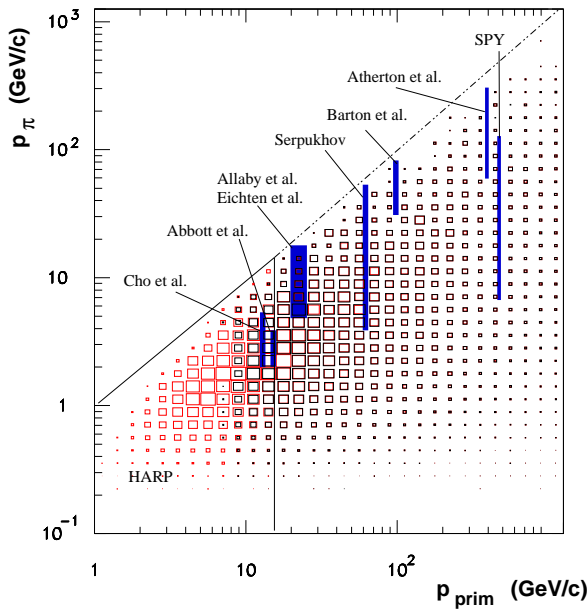,width=7.7cm}}
\caption{Summary of existing measurements of hadron
production as a function of primary momentum $p_{\rm prim}$ 
and secondary pion momentum $p_\pi$.
The black (red) boxes indicate the importance of each phase 
space region for downward (upward) atmospheric neutrino flux at
SuperKamiokande.  The difference is due to the Earth's magnetic field.
The vertical bars indicate where experimental data exist on low-$A$
nuclei. The indicated measurements are 
Abbott et al.~\protect\cite{Abbott92},
Cho et al.~\protect\cite{Cho71a},
Eichten et al.~\protect\cite{Eichten72},
Allaby et al.~\protect\cite{Allaby70},
Barton et al.~\protect\cite{Barton83},
Serpukhov~\protect\cite{Bozhko79a},
Atherton et al.~\protect\cite{Atherton:1980vj}, and
SPY~\protect\cite{Ambrosini:1998et}.
}
\label{fig:barbara}
\end{figure}

Figure~\ref{fig:barbara} shows the current status of hadron production
measurements for proton projectiles in the region below 1\,TeV primary energy.  The vertical
lines indicate the locations as a function of primary energy $E_p$ and
secondary energy $E$ where {\it some} measurements have been made.  An
additional degree of freedom, not indicated on the plot is the
transverse momentum $p_T$ of the secondary particles (the production
is symmetric in the $\varphi$ angle).  For most of the applications
described above, the total production as a function of $E_p$ and $E$
is of main interest (i.e. integrated over $p_T$). Given the limited
number of data points this integration is ambiguous and depends on the
functional form used for the extrapolation to unmeasured phase space
regions (for example, see discussion in \cite{Engel:1999zq}).

Since the $p_T$
distributions control the lateral extent of a shower, the functional
form of the hadron yields as a function of $p_T$ is also important for
EAS simulations. 
Atmospheric neutrinos are only observed one at a time (i.e. the
chance of detecting two neutrinos from the same cosmic ray are
negligible), the form of the $p_T$ distribution is therefore
less important for atmospheric neutrinos, however it is important to have
the yield integrated over the full region of $p_T$.
Neutrino beams select and focus pions from specific
regions of phase space and so $p_T$ distributions are important here
also.  

Many of the measurements shown in figure~\ref{fig:barbara} have been
made using `single armed spectrometers'.  These are devices where the
secondary particles are measured with a string of magnets tuned to
transmit particles of a given momentum at a time.  Examples of
experiments of this type are given in
Refs.~\cite{Eichten72,Atherton:1980vj,Ambrosini:1998et}. 
The experiments can be `ready
made' by using the magnets in a secondary beam at an accelerator
center such as CERN as the measuring device.  A big advantage of this
technique is that precise particle identification tuned to the
momentum which the spectrometer is being used can be made to separate
$\pi$, K and p extremely well. The measurements are somewhat
painstaking, since only one angle and momentum of the secondary
particles can be measured at a time and this is the reason why these
measurements do not yet cover sufficient points in phase space for
building models of hadron production.

A second technique is to use an emulsion stack in an accelerator beam.
These experiments are limited in statistics and particle
identification, however they can measure the complete phase space
region in one exposure.  Bubble chamber experiments have also been
used, e.g.~\cite{Mueck:1972qz}.

The experiments described below aim to measure a large fraction of (or
the complete) secondary phase space in one setting by using large
acceptance detectors.  This will allow complete $p_T$ distributions to
be obtained in large regions of $E$.  The experiments are not quite so
good at particle identification as the single-arm experiments.


\subsection*{HARP}

\begin{figure}[htb!]
\begin{center}
\includegraphics*[width=7.6cm]{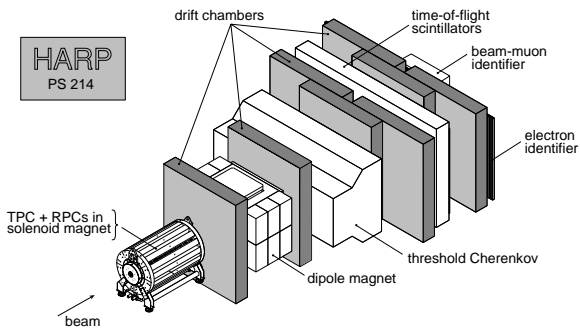}
\end{center}
\caption{Layout of the HARP experiment
\protect\cite{Catanesi:1999xr,Ellis:2003tt}.}
\label{fig:harpexp}
\end{figure}

The HARP hadron production experiment~\cite{Catanesi:1999xr} was built
specifically to address the lack of coverage of phase space with good
hadron production data for the applications as highlighted at the
beginning of this article.  A comprehensive range of targets (LH$_2$,
LD$_2$, Be, C, LN$_2$, LO$_2$, Al, Cu, Sn, Ta, Pb, H$_2$O) of
different atomic weights $A$ and thickness have been exposed to beams
from the CERN proton synchrotron (PS) with momenta between 1.5~and
15~GeV.  Beams of both positive and negative targets have been used.
Several replicas of neutrino beam targets have also been exposed.
There are several solid targets with $A$ close to that of air and
cryogenic (liquid) nitrogen and oxygen targets have also been exposed.

The beam is produced from a primary PS target at 24~GeV and the
secondary particles are momentum selected and tagged with gas Cherenkov
and time of flight detectors in the beam.  All interactions are
triggered and the type of beam particle determined offline; thereby,
interactions with parent protons, kaons and pions are all available.
The beam instrumentation also measures the trajectory of the incoming
particle so that the impact position on the target can be
reconstructed.

The detector is shown in figure~\ref{fig:harpexp}.  There are two
distinct groups of detector to select different regions of phase
space; large angle and small angle with respect to the beam direction.

The large angle detectors consist of a 80\,cm diameter by 1.5\,m
long time projection chamber (TPC) enclosed in a 0.5~T
solenoidal magnetic field.  A hole in the center of the TPC allows the
target to be inserted in the center of the sensitive region of the
detector so that backward going particles can be measured.  The TPC is
the primary detection device for large angle tracks and pulse height
information from the detector allows particle identification through
ionization loss (d$E$/d$x$) measurement \cite{Prior:2003ci}.
A layer of resistive plate
detectors (RPCs) with $\sim160$~ps resolution surrounds the detector
for particle identification by time of flight \cite{Bogomilov:2003br}.
The trigger is
provided by a series of scintillating fiber detectors mounted along
the central hole of the TPC.

The tracking detectors in the forward region (small angle) are drift
chambers with three views (vertical and $\pm5^\circ$).
A vertical magnetic field bends the
particles in the horizontal plane to determine the particle momentum.
Downstream, a plane of scintillation counters with $\sim200$\,ps timing
resolution is used to identify the particles by time of flight and
provides $3\sigma$ $\pi/$p separation up to 4.5\,GeV.  A large
(31\,m$^3$) Cherenkov detector filled with C$_4$F$_{10}$ provides good
$\pi/$p separation at high momentum.  Electron identification is
provided by a scintillating fiber and lead sampling calorimeter.

\begin{figure}[hbt]
\centerline{\epsfig{file=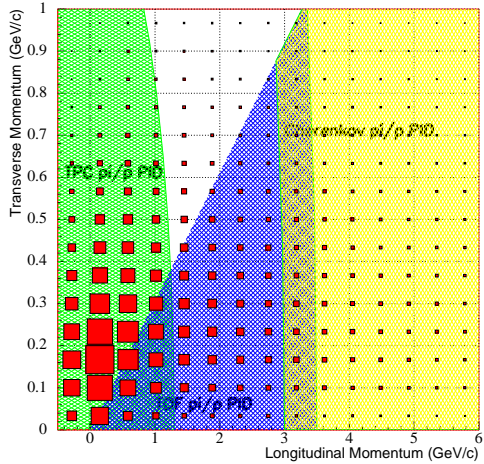,width=7cm}}
\caption{Secondary particle distribution in p-Be interactions at 15 GeV.
Phase space regions for different methods of particle identification are
superimposed.}
\label{HARP-PID}
\end{figure}

The different phase space regions with the corresponding
detector components employed for particle identification
are shown in Fig.~\ref{HARP-PID}. The squares show the expected secondary
particle distribution for 15 GeV protons on Be.

\begin{figure}[hbt]
\begin{center}
\epsfig{file=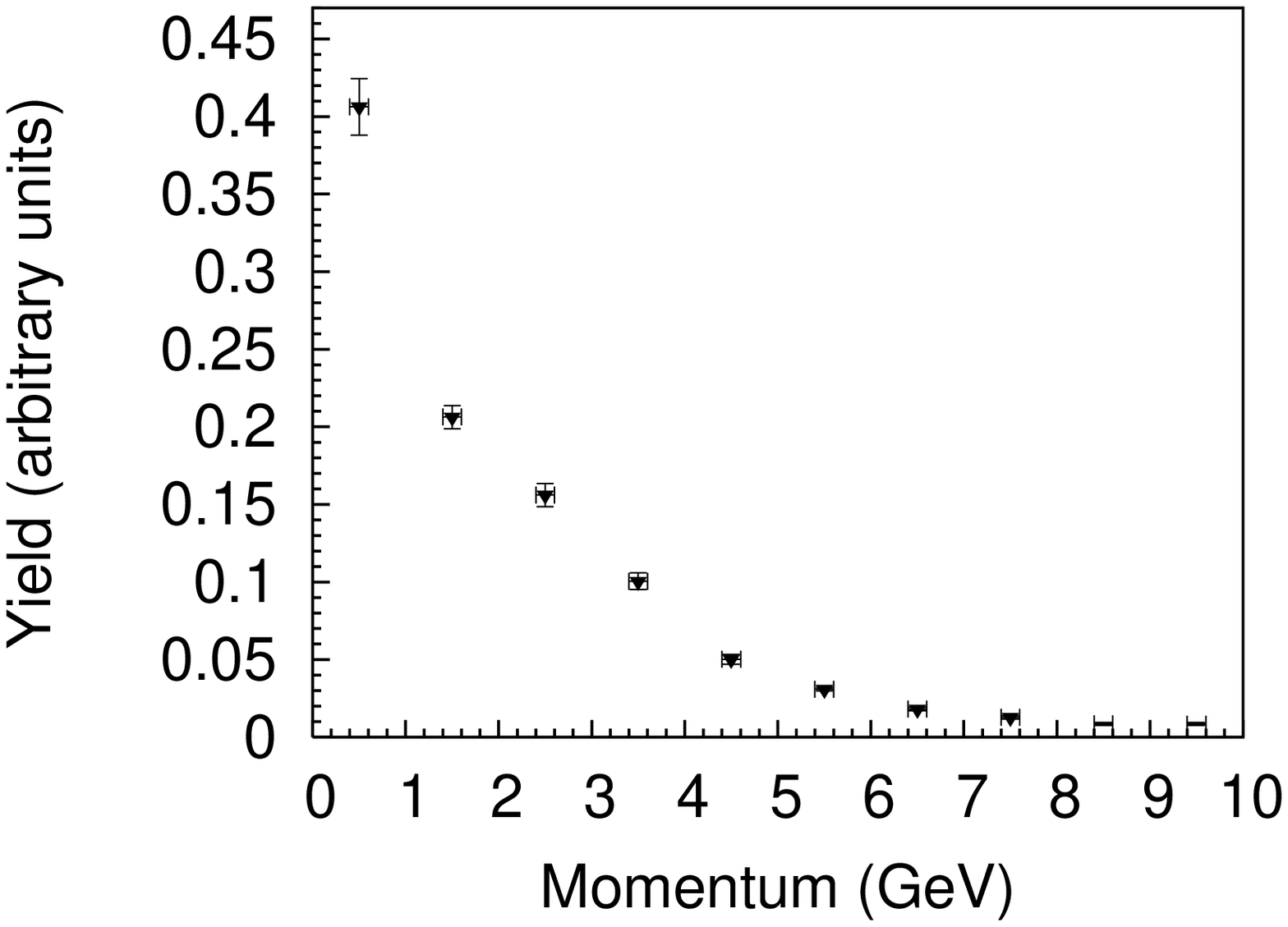,width=7cm}
\epsfig{file=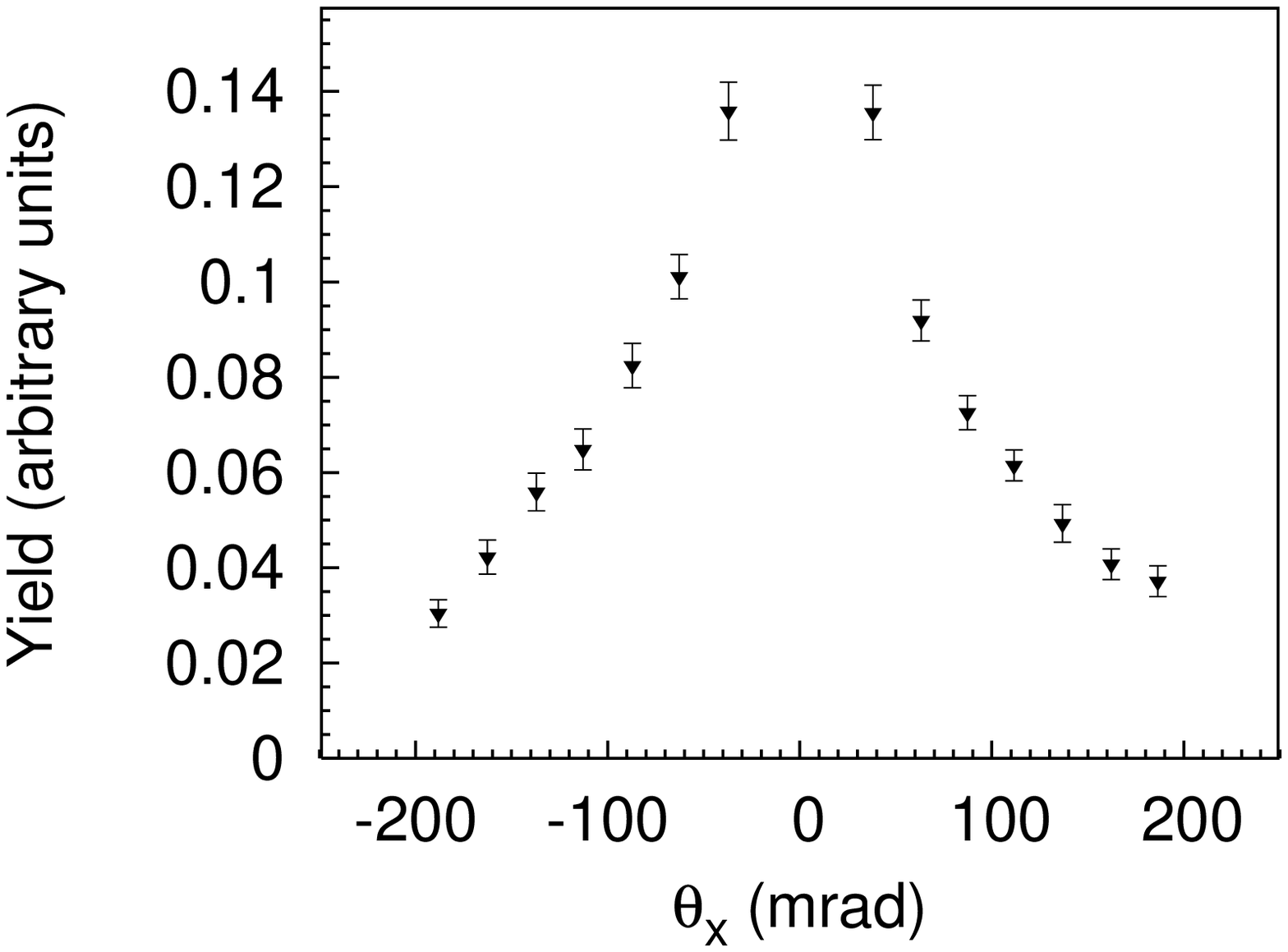,width=7cm}
\end{center}
\caption{Recent HARP hadron production data on a thin aluminium target
(K2K target replica) as a function of momentum and of angle in 
the horizontal plane \protect\cite{Kato04a}.}
\label{fig:harpresult}
\end{figure}

Preliminary data on a thin 20~mm (5\% of an interaction length)
aluminium target, a short section of a replica target for the K2K
experiment have become available recently~\cite{Kato04a}.
Figure~\ref{fig:harpresult} shows the yield of secondary particles 
identified as
pions as a function of both the momentum $p$ and opening angle
$\theta_z$ in the horizontal (bending) plane for momenta above
0.2~GeV/$c$.  Fiducial cuts were applied at $25<|\theta_z|<200$~mrad
and $|\theta_y|<50$~mrad, where $\theta_y$ is the opening angle in the
vertical plane.

\subsection*{NA49}

\begin{figure*}[hbt]
\begin{center}
\epsfig{file=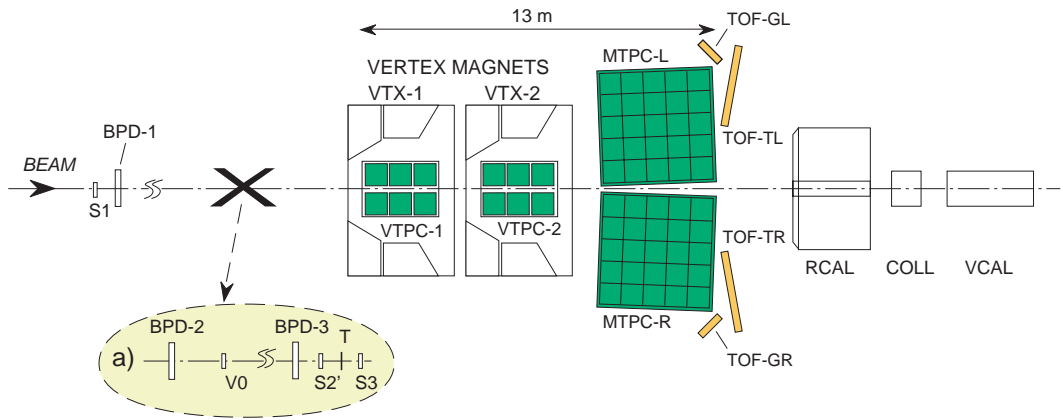,width=14cm}
\end{center}
\caption{Plan layout of the NA49 experiment.  The four TPCs are shown
(VTPC1,2 with magnetic field and MTPC L,R).  The position of the target 
is indicated by a 'T' in the inset a) immediately in front of VTPC1.  
Beam position detectors (BPD) and trigger counters S1,2,3 are also shown.}
\label{fig:na49}
\end{figure*}

The NA49 experiment~\cite{Afanasev:1999iu} 
was first commissioned in 1994 and has
been used since then with both lead ion and proton beams from the CERN
SPS.  The experiment is designed to cope with the extremely high
multiplicity environment of a Pb--Pb collision at 400~GeV/$Z$.  

NA49 is installed in the H2 beam line at the CERN SPS.  The experiment
(Fig.~\ref{fig:na49}) 
consists of two (2\,m$\times$2.5\,m$\times$0.98\,m overall dimensions)
TPCs (VTPC1 and VTPC2) supported inside 1.7~T super-conducting magnets.  Downstream,
there are two further TPCs (3.9\,m$\times$3.9\,m$\times$1.8\,m) (MTPCL and MTPCR) which
measure the tracks once they emerge from the magnetic field.
Approximately 100 samplings of the ionization loss are made per track
in the combined TPCs to determine
the particle identification from the relativistic rise section of the
energy loss curves.  This is the principle form of particle
identification in the experiment.  The target is mounted in front of
the first TPC, so, unlike HARP, only forward tracks may be measured.
The detector has excellent acceptance for particles with
$x_f>0$.  Particles with momenta above 80~GeV/$c$, pass
through the gap between TPCs (essential for heavy ion
running to avoid space charge from ionization from the
uninteracted beam).  A new thin gap-TPC was added to the
experiment to allow track detection (but not particle ID)
in this region.

A differential beam Cherenkov detector (two compartments, with pressures set
appropriately to be just above and below the Cherenkov threshold of the
desired particle) is used in the trigger so that only events which
originated from protons are in the data sample.  A beam particle
tracking system (BPD) allows the impact position on the target to be known
precisely.

The
experiment has recently been run with a low-bias trigger with a carbon
target for the purpose of measuring hadron production for atmospheric
neutrinos.  A one week run with protons at 158\,GeV and 100\,GeV has
yielded a sample of data with over 500,000 tracks \cite{Barr:2003ia}
and the analysis will be completed shortly.

The NA49 experiment has finished its data taking programme.
Currently a new collaboration is forming to perform an extended series of
measurements~\cite{NA49-SPSC-proposal}, including hadron production
measurements at CERN.  The experiment will use the existing NA49
apparatus and it is intended to collect a considerably more extensive
set of measurements than obtained in the pilot experiment described
above. It is planned to perform a primary energy scan from 
40 to about 200 GeV and a variety of
projectiles will be studied including protons and light nuclei such as 
helium and carbon.


\subsection*{Summary}

There are many reasons why the world's hadron production data is not
enough.  In particular series of many points in phase space are
required to allow the functional form of hadron production
measurements to be known precisely across all $p_T$ and secondary
particle momenta.  Two new measurements at CERN, HARP (a purpose built
experiment) and NA49 have been described.  Data at both experiments
have been collected and are currently being analyzed.  Preliminary
results from HARP have recently become available.  

Additionally,
there are new experiments in the USA which have, or soon will have
hadron production data.  The E910 experiment at Brookhaven~\cite{E910}
has data in the range 6--18\,GeV on beryllium \cite{Chemakin02a,Link04a}
and the MIPP experiment (specifically
built for hadron production measurements) at FNAL~\cite{MIPP} has
concluded its engineering run in 2004 and is
scheduled to collect data in the primary energy range 5--120 GeV in 2005
\cite{Raja:2005sh}.

\subsection*{Acknowledgements}

The authors would like to thank members of both the HARP and NA49
collaborations for help in the preparation of this contribution and
in particular G.~Catanesi and J.~Panman for their help during
preparation of the presentation. They would also like to thank
T.~Gaisser, S.~Robbins, and T.~Stanev for many useful discussions
and help on simulation of cosmic rays.  One of the authors, R.E.,
gratefully acknowledges the collaboration with A.~Haungs, D.~Heck,
C.~Meurer, and M.~Roth on the importance of low-energy interactions
in air showers.


\end{document}